\documentclass[superscriptaddress,twocolumn]{revtex4}
\pdfoutput=1
\usepackage{graphicx,color,url}
\definecolor{b}{rgb}{0,0,1.0}
\definecolor{r}{rgb}{1,0,0}
\definecolor{g}{rgb}{0,1,0}

\begin{document}

\newcommand{\SZFKI}{Research Institute for Solid State Physics and Optics,
 P.O. Box 49, H-1525 Budapest, Hungary}

\newcommand{\HASBUTE}{HAS-BUTE Condensed Matter Research Group, 
Budapest University of Technology and Economics, H-1111 Budapest, Hungary}

\title{Shear Zone Refraction and Deflection in Layered Granular Materials}

\author{Tam\'as B\"orzs\"onyi}

\email{btamas@szfki.hu}
\affiliation{\SZFKI}
\author{Tam\'as Unger}
\affiliation{\HASBUTE}
\author{Bal\'azs Szab\'o}
\affiliation{\SZFKI}

\begin{abstract}
Refraction and deflection of shear zones in layered granular materials was 
studied experimentally and numerically. We show, that (i) according to a recent
theoretical prediction [T. Unger, Phys. Rev.  Lett. {\bf 98}, 018301 (2007)] 
shear zones refract in layered systems in analogy with light refraction, 
(ii) zone refraction obeys Snell's law  known from geometric optics and (iii)
under natural pressure conditions (i.e.  in the presence of gravity) the zone
can also be deflected by the interface so that the deformation of the high
friction material is avoided. 
\end{abstract}

\maketitle

When granular materials deform under external stress the
deformation is often localized into narrow regions. These shear zones
\cite{fehe2003,feme2004,unto2004,lu2004,chle2006,feme2006,desa2006,dele2007,toun2007,un2007,un2007NP,riun2007,safe2008,jo2008,ja2008,hifa2008,mude2000,gdrmidi2004}
act as internal slip surfaces between solid-like blocks of the bulk.  The
formation of shear zones is a crucial deformation mechanism in fine
powders, sand and soil (landslides). Geological faults are themselves
large scale examples of shear zones. Here we study experimentally a
recent theoretical prediction that shear zones behave in striking analogy
with geometric optics \cite{un2007,un2007NP}, and compare our results
with numerical simulations. We show that shear zones alter their
orientations when crossing media boundaries similarly to light refraction
but here the frictional properties of the materials take the role of the
optical refractive index.  We find that the refraction phenomenon also
exists in the presence of gravity, i.e. under natural pressure
conditions. In certain configurations we observe another effect, namely
that shear zones can be deflected by the material interface.

In the present experiments we use two materials with different frictional
properties. One material is corundum which consists of angular grains 
\begin{figure}[htb]
\includegraphics[width=\columnwidth]{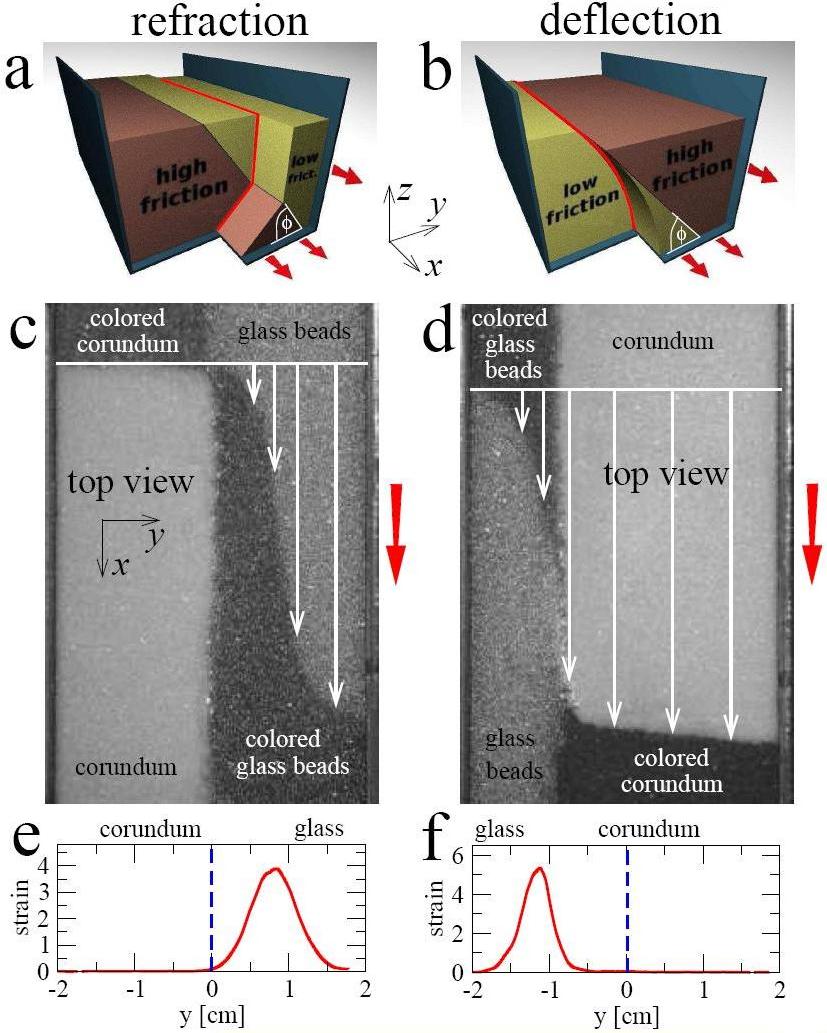}
\caption{(color online) Schematic illustration of (a) refraction and (b) 
deflection of a shear zone in layered granular systems according to
our experimental setup. 
(c)-(d) Top view of the corresponding experiments after deformation. 
The measurement was done with colored and uncolored particles in order to
visualize the displacement profile (illustrated by white arrows). 
(e)-(f) Shear strain at the top measured during the experiment. The dashed
line shows the position of the split line at the bottom.}
  \label{setup}
\end{figure}
therefore it has higher effective internal friction than the other material
consisting of glass beads. To characterize the difference in the internal 
frictions the angles of repose ($\theta_r$) were determined by the method 
used in \cite{boha2008} and were $\theta_r^{gla}=21.9^\circ$ for glass beads
while $\theta_r^{cor} = 33.2^\circ$ for corundum which is somewhat higher than
the typical value for sand ($\theta_r^{san} = 30.5^\circ$). 
The ratio $\tan\theta_r^{cor}/\tan\theta_r^{gla} = 1.63$ gives a reasonable
contrast for the internal friction. 
The interface between the two materials (illustrated with different 
colors in Figs.~\ref{setup}(a) and ~\ref{setup}(b)) is a tilted plane.  
The inclination of this tilted plane with respect to horizontal defines the 
interface angle $\phi$. Shearing was
performed in a straight split bottom cell \cite{desa2006,dele2007,riun2007,ja2008} 
with internal cross section of $4.2$ cm x $4.5$ cm.
The shear cell included two 60 cm long L shaped sliders (cell wall), one of 
which was slowly translated in the experiments according to the red (gray) 
arrows (with total displacement between $5$ and $6$ cm), which creates the shear 
zone indicated by the red line on the sketch. The presence of the 
interface leads to refraction or deflection of the shear zone depending on the 
configuration. This is in contrast to the simple case when the cell is 
filled with one material and the shear zone is formed along a vertical plane
\cite{riun2007,dele2007,desa2006}.

The idea of refraction for shear zones was proposed earlier \cite{un2007}
and studied numerically in a special geometry with cylindrical walls and
periodic boundary conditions. In that setup both ends of the shear zone
were fixed and the system was sheared under constant pressure in zero
gravity. These computer simulations showed that the zone changes its
orientation at the interface and the extent of refraction follows Snell's
law of light refraction. With the present work we were able to
reproduce the effect of refraction experimentally, moreover we show that
the phenomenon exists also under more realistic circumstances. In our case
the position of the shear zone is free at the top surface and gravity keeps
the grains together leading to a natural pressure gradient. 
This geometry allowed us also to detect the effect of deflection where the
interface modifies the position of the shear zone in such a way that the
deformation of the high friction region is completely avoided. Deflection
was not observed in the original cylindrical cell \cite{un2007} because the
two ends of the zone were fixed on different sides of the interface.
Therefore, the zone was forced to cross the interface and to enter the high 
friction material.

The deformation in our case was measured as follows. For both materials we
used two samples with different colors, therefore four regions appear at
the top surface as it is seen in Figs.~\ref{setup}(c) and \ref{setup}(d) .  
\begin{figure}[ht]
\includegraphics[width=\columnwidth]{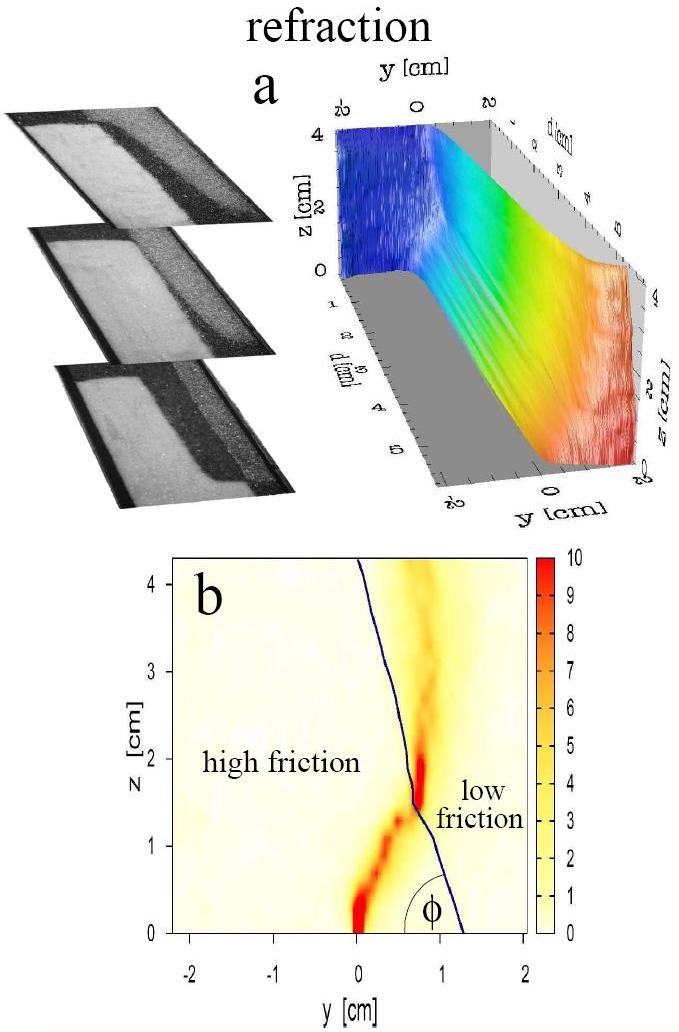}
\caption{(color online) (a) The reconstructed displacement profiles in the 
  bulk for refraction for $\phi=74^\circ$ taken by carefully removing the 
  material layer by layer (sample images taken during excavation are shown
  on the left hand side). Color (grayscale) corresponds to the deformation
  $d(y,z)$.  Movies demonstrating the excavation process can be seen 
  at \cite{movie}. (b) The corresponding shear strain distribution in the
  $y-z$ plane. The colors show the dimensionless shear strain $\gamma$ as
  defined in the text. The tilted blue (dark gray) line indicates the 
  position of the interface.}
  \label{expres1}
\end{figure}
Two of these regions correspond to glass beads 
(grain size $d_p=0.56\pm0.02$\,mm for the colored sample 
and $d_p=0.48\pm 0.02$\,mm for the uncolored sample) 
while the other two correspond to corundum 
(grain size $d_p=0.33\pm0.02$\,mm for the colored sample
and $d_p=0.23\pm 0.02$\,mm for the uncolored sample). Using particles 
with different sizes allowed the separation of samples after each
experiment, but have been shown not to affect the results \cite{movie}.
This way not only the interface of the two materials is visualized 
(oriented vertically in Figs.~\ref{setup}(c) and \ref{setup}(d)) but the 
displacement profile is also
directly seen, as it is illustrated with white arrows in
Figs.~\ref{setup}(c) and \ref{setup}(d). 
The deformation of the surface layer was recorded during translation by a 
video camera, movies recorded during the experiments can be seen at \cite{movie}.
The displacement at the top surface as a function of the position $y$ was 
determined by digital image analysis. The corresponding
shear strains (see Figs.~\ref{setup}(e) and \ref{setup}(f)) show that the shear
zones are shifted away from the split at the base of the cell. 
They are not in the middle (at $y=0$) where they
would be in the absence of the interface i.e. for a homogeneous material.

\begin{figure}[ht]
\includegraphics[width=\columnwidth]{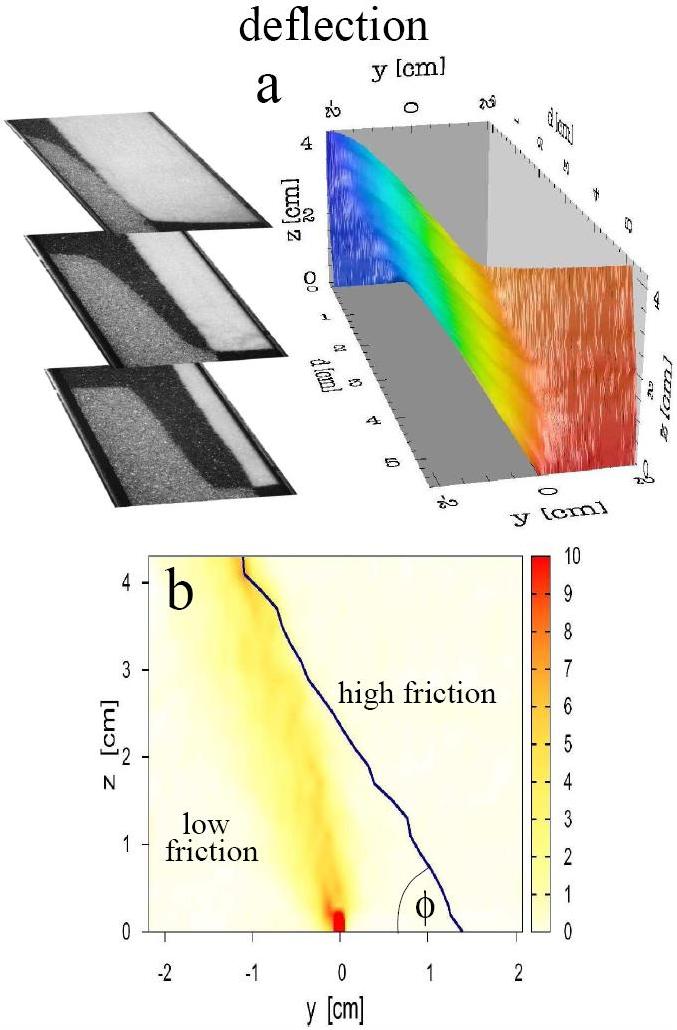}
\caption{(color online) (a) The reconstructed displacement profiles in the 
  bulk for deflection for $\phi=59^\circ$ taken by carefully removing the 
  material layer by layer (sample images taken during excavation are shown).
  Movies demonstrating the excavation process can be seen at 
  \cite{movie}.  (b) The distribution of the shear strain $\gamma$ in 
  the $y-z$ plane. The tilted blue (dark gray) line indicates the position of 
  the interface.}
  \label{expres2}
\end{figure}

After each experiment the displacement profile in the bulk $d(y,z)$ was
reconstructed by removing the top surface of the material carefully layer
by layer {\cite{feme2006}. We used a commercial vacuum cleaner with
additional extension tubes to provide a slow flow rate. The procedure 
takes several hours for one experiment. The displacement
  profiles are presented in Figs.~\ref{expres1}(a) and \ref{expres2}(a) for
  refraction and deflection, respectively together with sample images taken
  during the excavation process.  The gradient of the displacement provides
  the local shear strain $\gamma$ inside the bulk $\gamma= \frac{1}{2}
  \left[ \left(\partial_y d \right)^2+ \left( \partial_z d \right)^2
    \right]^{1/2}$. By plotting the shear strain $\gamma$ as the function
  of position (Figs.~\ref{expres1}(b) and \ref{expres2}(b)) the structure
  of the shear zone becomes visible.  In Fig.~\ref{expres1}(b) the zone,
  starting from the bottom, takes a short path towards the interface and by
  reaching the low friction region it changes direction abruptly and heads
  straight to the top. In Fig.~\ref{expres2}(b) where the zone starts in
  the low friction material it deflects to avoid the high friction region
  even if it takes a much longer path.

The behavior can be well understood qualitatively based on a simple
variational model of shear zones \cite{unto2004,un2007}. This model regards
the shear zone as an infinitely thin slip surface and states that the shape
of the slip surface is chosen according to minimum energy dissipation. For
our straight geometry this is reduced to the condition that the
minimization of the following integral provides the shape of the zone:
\begin{equation}
  \int \mu p \, \, \text{d}l = \text{min.} \, ,
\label{path}
\end{equation}
where the local effective friction coefficient $\mu$ times the local
\begin{figure}[ht]
\includegraphics[width=\columnwidth]{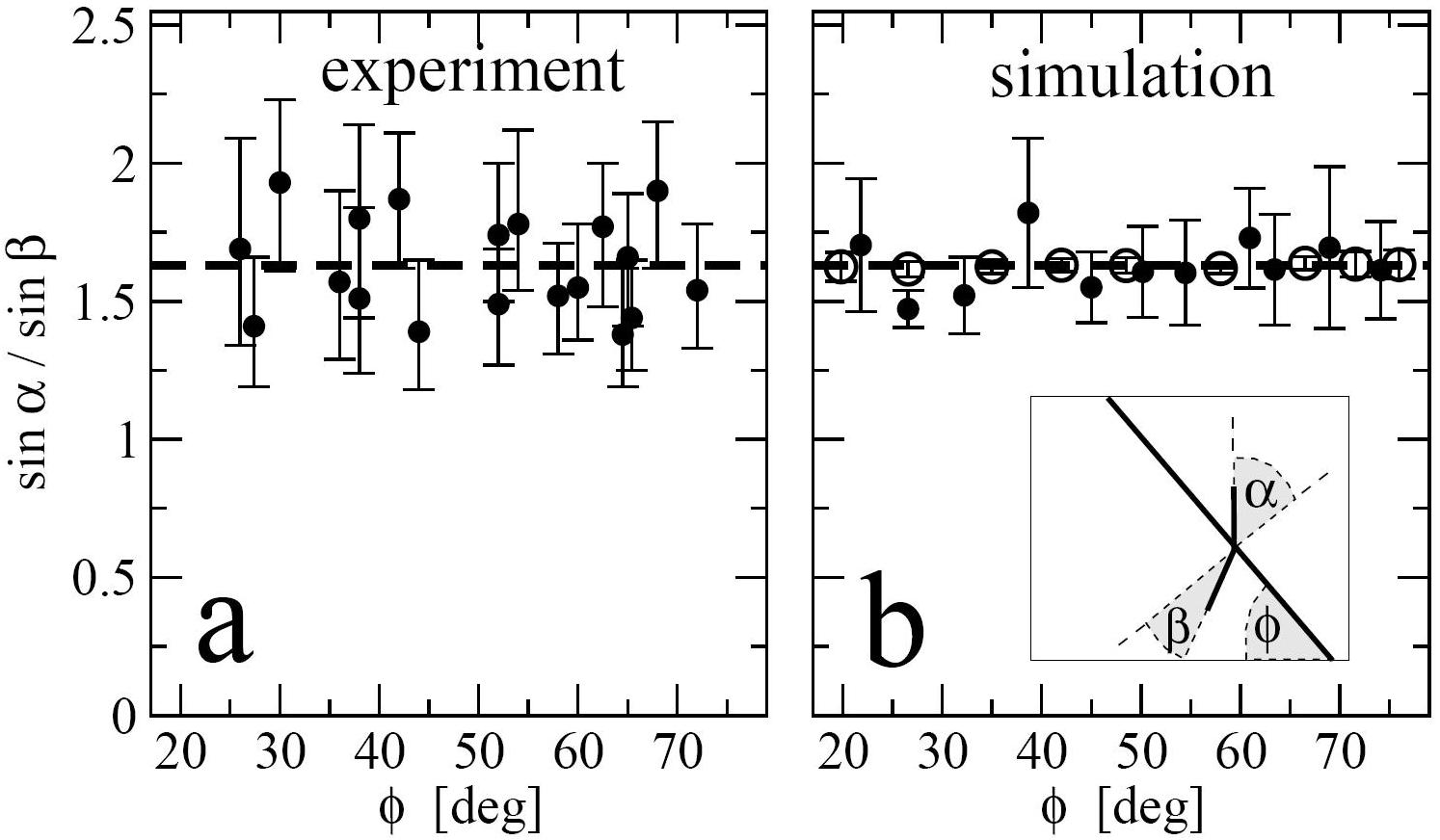}
\caption{The ratio $\sin\alpha/\sin\beta$ ($\alpha$ and $\beta$ being the 
  angles of incidence)
  for zone refraction as a function of the interface angle $\phi$
  with respect to horizontal obtained from (a) experiments and (b) numerical 
  simulations. For the case of simulations the open data points were taken in 
  the absence of gravity (i.e. homogeneous pressure).
  The horizontal dashed lines correspond to the ratio of the effective frictions 
  $\mu_{\text{cor}} / \mu_{\text{gla}}=1.63$ determined from the angles of 
  repose for the two materials used. The inset of (b) illustrates the angles 
   $\alpha$, $\beta$ and $\phi$.} 
  \label{expres3}
\end{figure}
pressure $p$ is integrated along the path that is taken by the shear zone
in the $y$-$z$-plane section. Thus not the length of the path but the
length weighted by $\mu p$ is minimized by the shear zone. This is quite
similar to Fermat's principle of optics where the length weighted by the
refraction index is minimized by light beams. This analogy leads to the
idea that similar refraction effects were expected in these two distant
fields of physics.

We performed experiments for numerous orientations of the interface 
to test whether zone
refraction obeys Snell's law known from optics, which  reads
$\sin\alpha/\sin\beta=\mu_{\text{cor}} / \mu_{\text{gla}}$ in the present
case.  Here $\alpha$ and $\beta$ are the angles of incidence (see inset
of Fig.~\ref{expres3}(b) for illustration) and the relative index of
refraction is replaced by the ratio of the effective frictions
$\mu_{\text{cor}} / \mu_{\text{gla}}$~\cite{un2007}.
Fig.~\ref{expres3}(a) shows $\sin\alpha/\sin\beta$ as a function of the
interface angle $\phi$ with respect to horizontal and the
value of $\mu_{\text{cor}} / \mu_{\text{gla}} = 1.63$ (dashed line)
that was determined experimentally based on the two angles of repose
\cite{boha2008}. As it is seen the data are independent of $\phi$ 
(within errors) for a wide range of $\phi$ and scatter around the 
expected value 1.63 in agreement with Snell's law.

In order to better compare the predictions of the above concept with our 
experimental data we performed computer simulations based on the
fluctuating narrow band model \cite{toun2007}. This method follows the
concept of minimum dissipation but also takes the fact into account that
the optimal slip surface has to be found in a random medium. 
Technically, in the simulation we search for an optimal path in a
two dimensional underlying lattice which represents the cross section of the
shear cell. Statistical fluctuations are introduced by random strength of
the lattice-bonds. Depending on
the random realization of the medium different slip surfaces are
obtained. Taking ensemble average over random realizations one arrives at a
continuous displacement profile instead of a thin slip surface. To
incorporate the ratio of the effective frictions $\mu_{\text{cor}} /
\mu_{\text{gla}}$ we use the same ratio for the average bond strength
between the low and the high friction regions. There is one fit parameter
in the simulation, the extent of fluctuations, which is chosen so that
the width of the shear zone at the free surface approximately matches
the experimental value. More
details on this method can be found in \cite{toun2007}.

Numerical simulations (Fig.~\ref{expres3}(b)) confirm with higher
accuracy that $\sin\alpha/\sin\beta$ matches the expected value of
$\mu_{\text{cor}} / \mu_{\text{gla}} = 1.63$.  Simulations also revealed,
that this is true for both the absence and the presence of
gravity where the pressure was taken to be constant and proportional to
the depth, respectively \cite{riun2007}. In gravity the shear zone is
slightly curved which is a direct consequence of the pressure gradient
and makes the measurement of refraction more difficult. Under homogeneous
pressure where the curvature is absent, the accuracy of data is much
higher (see open circles in Fig.~\ref{expres3}(b)).  

For a more detailed quantitative comparison two simulations are shown
in Fig.~\ref{simres} which correspond to the experimental configurations
presented in Figs.~\ref{expres1} and \ref{expres2}. The shear strain 
distributions obtained by
\begin{figure}[ht]
\includegraphics[width=\columnwidth]{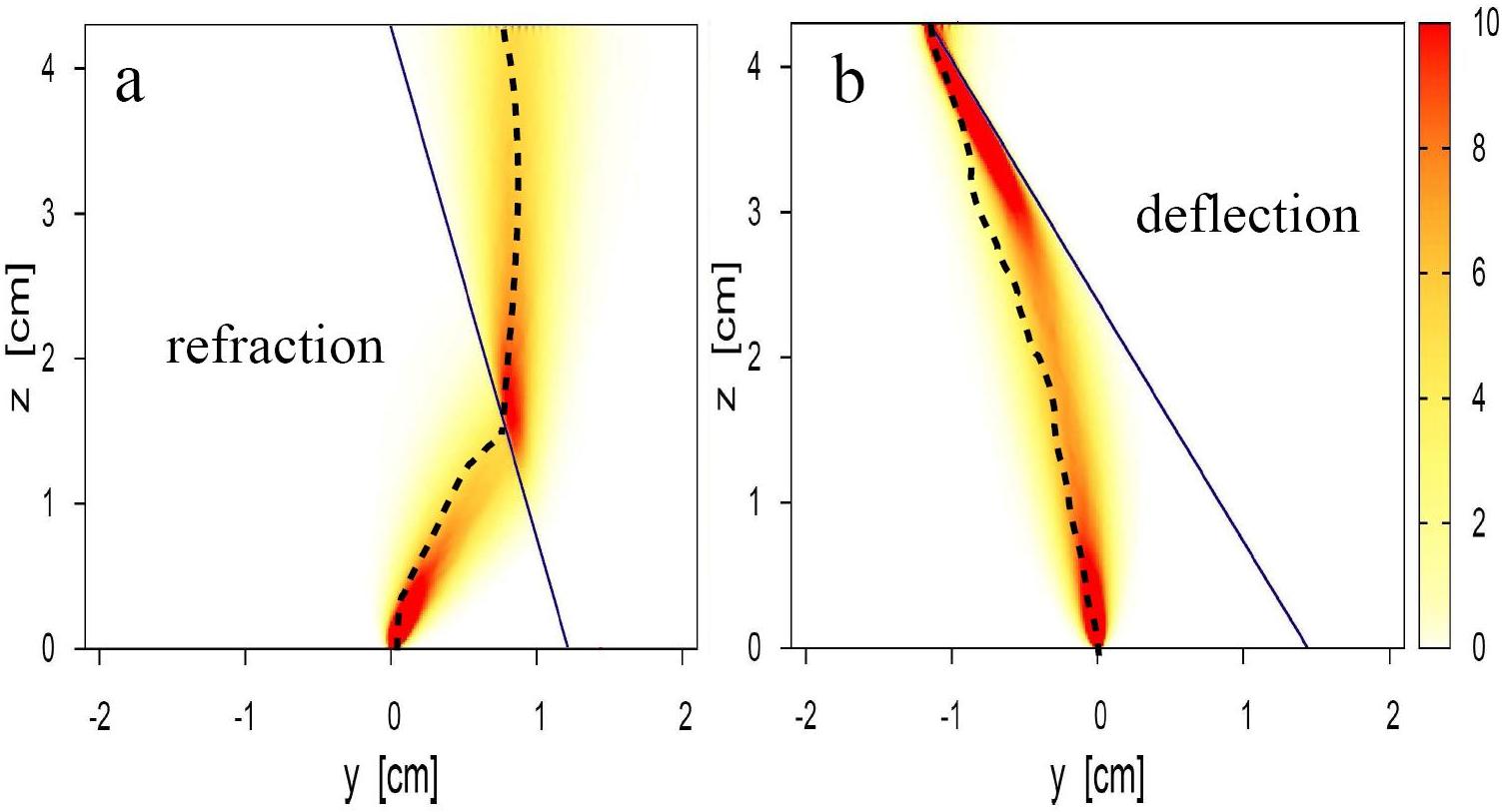}
\caption{(color online) Distributions of the shear strain $\gamma$ obtained
in numerical simulations for (a) refraction and (b) deflection based on the
fluctuating narrow band model. The dashed lines show the center of the
experimentally achieved shear zones, for comparison.}
\label{simres}
\end{figure}
the simulation are in very good agreement with the experiments. The shape
of the shear zones and other details are nicely reproduced, e.g. the
largest values of the shear strain. In the case of refraction the largest
strain is obtained at the bottom and right above the interface near to the
refraction point, while for deflection the high shear strain is located at
the bottom and at the top of the material.  This latter effect leads to a
narrower shear zone at the top in our second configuration (deflection)
than in the first one (refraction), which is also nicely detected in the
experiments - compare the width of the zone in Figs.~\ref{setup}(e) and
\ref{setup}(f).  Comparing the center positions of the shear zones at the top
gives also nice agreement. We obtain $y=0.85$ cm in the simulation and
$0.81$ cm experimentally in case of refraction, while for deflection the
positions are $-1.11$ cm and $-1.15$ cm in the simulation and experiment,
respectively.

In this work we studied shear zones in granular systems where two
different materials are layered on top of each other. On one hand we
provided an experimental evidence of the phenomenon of zone refraction,
on the other hand we found a new effect where the shear zone is deflected
by the interface. For the case of zone refraction a close analogy can be 
drawn with light refraction in optics. Our experimental and numerical data 
show, that the well known Snell's law is valid for this granular system.
The presented experiments and simulations clearly
demonstrate that both effects - refraction and deflection - alter
significantly the position of the shear zone, thus they may play 
important role in various industrial and civil engineering
applications. Moreover, we expect that our findings may be relevant in
geological processes where the materials involved (soil, rock) have
naturally layered structures.

Further experiments in layered materials are on the way to explore the
bulk deformation with an independent method (MRI imaging).
We recently learned about a parallel experimental
study on shear zone refraction obtained in a vertical cylindrical cell
\cite{knbe2009}.

\noindent \textbf{Acknowledgments}

\noindent
The authors are thankful for discussions with J\'anos Kert\'esz and Ralf Stannarius.
T.B. and T.U. acknowledge support of the Bolyai J\'anos 
research program, and the Hungarian Scientific Research Fund 
(Contract Nos.\ OTKA F060157 and PD073172).

\end{document}